\documentclass[12pt]{article}

\usepackage{epsfig,bm,titlesec}
\usepackage{hyperref}
\usepackage{graphicx}

\begin{document}

\title{Charge, strangeness and radius of strangelets}

\author{
 X.\ J.\ Wen,$^{1,2,3}$\ \
 G.\ X.\ Peng,$^{1,2,3}$\footnote{E-mail address: gxpeng@ihep.ac.cn}\ \
 Y.\ D.\ Chen,$^{2,3}$\\
 \small\sl
 $^1$China Center of Advanced Science and Technology (World Lab.),
 Beijing 100080, China\\[-0.5cm]
\small\sl
 $^2$Institute of High Energy Physics, Chinese
     Academy of Sciences, Beijing 100049, China\\[-0.5cm]
 \small\sl
 $^3$Graduate University of Chinese Academy of Sciences,
     Beijing 100049, China
      }

\maketitle

\begin{abstract}
We investigate, at both zero and finite temperature, the properties
of strangelets versus the electric charge $Z$ and strangeness $S$.
The strangelet radius is not a monotonic function of either charge
or strangeness, and a minimum is reached in the ($Z$, $S$) plane.
However, the thermodynamically stable strangelets do not correspond
to the radius minimum. The minimum radius always appears at positive
strangeness, while the stable radius may appear at negative
strangeness for very small baryon numbers. For large baryon numbers,
the stable radius is proportional to the cubic root of baryon
numbers, but inversely proportional to the square root of the
confinement parameter in the present model. If bulk strange quark
matter is absolutely stable, the reduced size of strangelets is
about 1 fm, which may be relevant for the analysis of the strangelet
production and detection.
\end{abstract}


\section{Introduction}

Strange quark matter (SQM) has been one of the hot topics in nuclear
physics since Witten's conjecture \cite{Witten1984} that SQM might
be the true ground state of strong interactions. Lumps of SQM are
customarily called strangelets \cite{Farhi84prd}, or for short,
slets \cite{Peng06plb}. There are basically two kinds of slets: one
is the ordinary slets without pairing
\cite{Farhi84prd,Berger1987,He1996a,Schaffner1997,Madsen00prl,Zhang2001},
the other is the color superconductivity slets
\cite{Madsen01prl,Peng06plb,Alford2006prd73}. The possible
terrestrial production of slets has been studied in high energy
heavy ion experiments, e.g., at the CERN SPS energies
\cite{Lourenco2002}. For a review on recent experimental searches of
quark-gluon plasma at Brookhaven RHIC, see
Ref.~\cite{Weiner2005ijmpe}. At high densities, quark matter may be
the most favorite phase, and could thus exist in the core of compact
stars \cite{Peng2005epl}. Strange star collisions could release
slets as part of the energetic cosmic rays, and some of the cosmic
ray slets might be on the way to our Earth
\cite{Finch2006jpg32,Monreal07jhep}. The Alpha Magnetic Spectrometer
(AMS-02) \cite{Sandweiss2004}, which is planned to operate on the
International Space Station, may offer the opportunity to detect
these slets \cite{Cheng2006}.

It is still an interesting open problem whether or not these cosmic
ray slets incident on top of the Earth's atmosphere can reach to the
ground or sea level. In fact, several exotic cosmic ray events with
anomalously low charge-to-mass ratio have been observed at rather
low altitudes \cite{Saito1990,Ichimura1993,Price1978}. In
literature, one finds a number of possible scenarios of slet
propagation which depends strongly on assumed size of slets. For
example, the radius of slets used in \cite{Wilk1996} is different
from that in \cite{Capdeville1995,Miyamura1995} by nearly an order
of magnitude. Since the mean free path of a slet in the atmosphere
is strongly dependent on its radius $R_{\mathrm{slet}}$, as, for
instance, in the simple law $ \lambda
=A_{\mathrm{air}}m_{\mathrm{N}}/[\pi(R_{\mathrm{slet}}+R_{\mathrm{air}})^2]
$ where $m_{\mathrm{N}}\approx 939$ MeV is the nucleon mass,
$A_{\mathrm{air}}\approx 14.5$ and $R_{\mathrm{air}} =1.12
A_{\mathrm{air}}^{1/3}\approx 2.73$ fm are, respectively, the mean
mass number and radius of the nucleus in the atmosphere, different
radii led to significantly different conclusions.

Meanwhile, the charge property of slets is very important, as shown
by Madsen \textit{et al.} who found that the slets of low
charge-to-mass ratio are favored in the ultra-high-energy cosmic
rays \cite{Madsen2003}, and the color flavor locked (CFL) slets have
charge $Z\approx$ 0.3$A^{1/3}$ where $A$ is the baryon number. Jaffe
\textit{et al.} also demonstrated that the slets with large baryon
number have positive charges $Z \sim A^{1/3}$ \cite{Berger1987}. In
addition to these positively charged slets, negative charges are
also possible for both ordinary \cite{Peng1999} and CFL slets
\cite{Peng06plb} in beta equilibrium. At the same time, the
strangeness fraction is also an important factor to determine the
stable configuration of slets \cite{Schaffner1997}.

Recently, we suggested a new quark mass scaling \cite{Wen2005} based
on the linear confinement, and the new model was applied to
investigating the properties of slets in full beta equilibrium. In
heavy ion collision experiments, however, the time scale is not
enough for perfect beta equilibrium. Moreover, the
charge/strangeness composition, and especially the size of slets,
are quite important and useful to analyse possibility of production
and detection of slets \cite{Wilk1996}. In this paper, therefore, we
study the relevant properties of slets with the new quark mass
scaling in \cite{Wen2005}, without imposing beta equilibrium, and
with focus on the slet size. It is found that the mechanically
stable radius of a slet with fixed baryon number and temperature is
not a monotonic function of either charge $Z$ or strangeness $S$.
The radius has a minimum in the $(Z,S)$ plane. However, the radius
minimum does not correspond to the minimum of free energy. We
determine the composition of thermodynamically stable slets by
minimizing the free energy. For a large baryon number $A$, the
corresponding slet radius is $R=[(3/4)^{1/3}/\sqrt{x_0D}]A^{1/3}$,
where $x_0=1.3278478$, and $D$ is the confinement parameter in the
present model. If SQM is absolutely stable, the reduced stable
radius is $r_{\mathrm{slet}}\equiv R/A^{1/3}\approx 1$ fm. In the
conventional bag model, slets always contain strange quarks. In the
present model, we find that very small slets tend to contain
anti-strangeness, and the ratio of charge to baryon number
increases.

This paper is organized as follow. In Sec.\ \ref{eos} we introduce
the thermodynamic treatment with density and temperature dependent
quark masses. The properties of strangelets related to the
strangeness and electric charge at both zero and finite
temperature are presented in Sec.\ \ref{strangelets}. A summary is
given in the final section \ref{sum}.

\section{Thermodynamics with density and temperature dependent quark masses}
\label{eos}

We start from the total free-particle thermodynamic potential density
\begin{equation} \label{Otmum}
\Omega=\sum\limits_i\Omega_i(T,\mu_i,m_i,R),
\end{equation}
where the summation index $i$ goes over $u, d, s$ quark flavors
, $T$ is the temperature, $m_i\ \mbox{and}\ \mu_i\ (i=u,d,s)$
are the corresponding quark masses and chemical potentials,
and $R$ is the slet radius. At finite
temperature, we treat the anti-quarks as a whole with quarks. The
contribution of the thermodynamic potential density from the density
of state $n_i^{\prime}(p,m_i,R)$ is given in the multi-expansion
approach \cite{Balian1970} as
\begin{eqnarray}
\Omega_i &=&
  -T\int_0^{\infty}
    \left\{
 \ln\Big[1+e^{-(\sqrt{p^2+m_i^2}-\mu_i)/T}\Big]
    \right. \nonumber\\
&& ~\left. \phantom{-T}
+\ln\Big[1+e^{-(\sqrt{p^2+m_i^2}+\mu_i)/T}\Big]
    \right\}
n_i^{\prime}(p,m_i,R) \mbox{d}p,
\end{eqnarray}
where the density of state is
\begin{equation}
n_i^{\prime}(p,m_i,R)
 =\frac{3}{\pi^2}
  \left\{
   p^2-\frac{3p}{2R}\mbox{arctan}\left(\frac{m_i}{p}\right)
   +\frac{1}{R^2}
    \left[
     1-\frac{3p}{2m_i}\mbox{arctan}\left(\frac{m_i}{p}\right)
    \right]
  \right\}.
\end{equation}
The three terms on the right are, respectively, the volume term,
surface term \cite{Farhi84prd,Berger1987} and curvature term
\cite{Madsen1993}.

To include the confinement interaction between quarks, we treat the
quark mass as density and temperature dependent, i.e.
$m_i=m_i(n_\mathrm{b},\mbox{T}),$\ where $n_{\mathrm{b}} = \sum_i
n_i/3$ with $n_i\ (i=u,d,s)$ being the quark number densities. This
means that the mass of quarks and antiquarks varies with state
parameters in a medium. We can divide the quark mass into two parts:
one is the current mass $m_{i0}$, the other is the interacting term
$m_{\mathrm{I}}$, i.e., $m_i=m_{i0}+m_\mathrm{I}$. In the present
calculations, we take the quark current masses $m_{u0} = 5$ MeV,
$m_{d0} = 10$ MeV and $m_{s0}$ = 120 MeV, respectively. Because the
strong interaction between quarks is a color interaction,
$m_{\mathrm{I}}$ is common for all quark flavors. The key point is
how to determine the interaction $m_{\mathrm{I}}$. In
Ref.~\cite{Zhang2002}, the ansatz $ m_{\mathrm{I}}
=\frac{B_0}{3n_{\mathrm{b}}}[1-(T/T_{\mathrm{c}})^2] $\ was
introduced. Because it caused an unreasonable temperature dependence
of the slet radius, another term linear in temperature was added
\cite{Zhang2001}. Based on the in-medium chiral condensates and
linear confinement, we recently derived a new quark mass scaling,
which can be expressed as \cite{Wen2005}
\begin{equation} \label{mass}
m_i =m_{i0}+\frac{D}{n_{\mathrm{b}}^z}
 \left[
  1-\frac{8T}{\lambda T_\mathrm{c}}
   \exp\left(-\lambda\frac{T_\mathrm{c}}{T}\right)
 \right],
\end{equation}
where $\lambda=\mbox{LambertW}(8)\approx 1.60581199632$ is a
constant, $T_\mathrm{c}$ = 170 MeV is the critical temperature. The
exponent $z$ was previously taken to be 1
\cite{Fowler1981,Chakrabarty1989,Benvenuto1995}. In order to be
consistent with the linear confinement, derivations based on the
in-medium chiral condensates \cite{Peng2006npa} showed that it is
more reasonable to take $z=1/3$
\cite{Peng2000prc61,Lugones2003ijmpd,Zheng2004}. The confinement
parameter $D$ can be constrained to a very narrow range by stability
arguments and we take $D^{1/2}=156$ MeV \cite{Peng2000prc62}.

The  particle number density for each quark flavor can be derived by
the following expression
\begin{eqnarray}\label{number}
n_i=-\frac{\partial\Omega}{\partial\mu_i} .
\end{eqnarray}
The pressure is
\begin{eqnarray} \label{Pexp}
P=\label{PMDTD}
  -\Omega-\frac{R}{3}\frac{\partial \Omega}{\partial R}
  +n_{\mathrm{b}}
     \sum_i\frac{\partial\Omega}{\partial m_i}
           \frac{\partial m_i}{\partial n_{\mathrm{b}}},
\end{eqnarray}
where the last term is due to the density dependence of quark masses
\cite{Peng2000prc62,Benvenuto1995}. The partial derivatives
$\partial m_i/\partial n_{\mathrm{b}}$ in Eq.~(\ref{PMDTD}) can be
easily obtained from the quark mass scaling in Eq.~(\ref{mass}),
i.e.,
\begin{eqnarray}
\frac{\partial m_i}{\partial n_{\mathrm{b}}} =
 -\frac{zD}{n_{\mathrm{b}}^{z+1}}
 \left[
  1-\frac{8T}{\lambda T_\mathrm{c}} \exp\left(-\lambda
  \frac{T_\mathrm{c}}{T}\right)
 \right]
 = -z \frac{m_\mathrm{I}}{n_\mathrm{b}}.
\end{eqnarray}
Accordingly the free energy density of the slets is
\begin{eqnarray} \label{Fexp}
F=\Omega -\sum_i \mu_i \frac{\partial \Omega}{\partial \mu_i}.
\end{eqnarray}

At zero temperature, the relevant integrations can be carried out.
The quark number densities in Eq.~(\ref{number}) become
\begin{eqnarray}
n_i
&=&
 \frac{\nu_i^3}{\pi^2}
 +\frac{9m_i^2}{4\pi^2R}
  \left[
   (x_i^2+1)\mbox{arctan}(x_i)
   -x_i\left(\frac{\pi}{2}x_i+1\right)
  \right]
\nonumber\\
&&
 +\frac{9m_i}{4\pi^2R^2}
  \left[
   (x_i^2+1)\mbox{arctan}(x_i)
   -x_i\left(\frac{3\pi}{2}x_i-1\right)
  \right].
\label{nzero}
\end{eqnarray}
where
$\nu_i=\sqrt{\mu_i^2-m_i^2}$ is the Fermi momentum of the quark flavor $i$
and $x_i\equiv \nu_i/m_i$.
The free energy density in Eq.~(\ref{Fexp}) becomes the energy density
\begin{eqnarray}
E
&=&
\sum_{i=u,d,s}
\frac{3m_i^4}{8\pi^2}
 \Bigg\{
\left[
 x\left(2x_i^2+1\right)\sqrt{x^2+1}-\ln\left(x+\sqrt{x^2+1}\right)
\right]
\nonumber\\
&&
+\frac{2}{m_iR}
 \left[
  \pi-x_i\sqrt{x_i^2+1} -\mbox{arcsh}(x_i)
  -2(x_i^2+1)^{3/2}\mbox{arccot}(x_i)
 \right]
\nonumber\\
&&
+\frac{2}{(m_iR)^2}
 \left[
  \pi+x_i\sqrt{x_i^2+1} +\mbox{arcsh}(x_i)
  -2(x_i^2+1)^{3/2}\mbox{arccot}(x_i)
 \right]
 \Bigg\}.
\label{Ezero}
\end{eqnarray}
And the pressure in Eq.~(\ref{Pexp}) becomes
\begin{eqnarray}
P
&=&
 \sum_{i=u,d,s}
  \frac{m_i^4}{8\pi^2}
  \Bigg\{
   x_i(2x_i^2-3)\sqrt{x_i^2+1}+3\mbox{arcsh}(x_i)
   -12z\frac{m_{\mathrm{I}}}{m_i}
    \left[
     x_i\sqrt{x_i^2+1}-\mbox{arcsh}(x_i)
    \right]
\nonumber\\
&&
+\frac{2}{m_iR}
 \left[
  3\pi\sqrt{x_i^2+1}-2\pi-4x_i\sqrt{x_i^2+1}
  +2\mbox{sh}^{-1}(x_i)
  -2(x_i^2+1)^{3/2}\mbox{arccot}(x_i)
 \right.
\nonumber\\
&& \phantom{+\frac{2}{mR}xx}
  \left.
-9z\frac{m_{\mathrm{I}}}{m_i}
\left(
 x_i\sqrt{x_i^2+1}+\pi-\pi\sqrt{x_i^2+1} -\mbox{sh}^{-1}\,x_i
\right)
\right]
\nonumber\\
&&
\frac{1}{(m_iR)^2}
\bigg[
 \pi(3\sqrt{x_i^2+1}-2)
 -2\mbox{arcsh}(x_i)
 -2(x_i^2+1)^{3/2}\mbox{arccot}(x_i)
\nonumber\\
&& \phantom{\frac{1}{(m_iR)^2}[}
\left.
-3z\frac{m_{\mathrm{I}}}{m_i}
\bigg(
 4\pi-3\pi\sqrt{x_i^2+1}
 +2x_i\sqrt{x_i^2+1}
\right.
\nonumber\\
&& \phantom{\frac{1}{(m_iR)^2}[-6z\frac{m_{\mathrm{I}}}{m_i}xx}
 +4\mbox{arcsh}(x_i) -2(x_i^2+1)^{3/2}\mbox{arccot}(x_i)
\bigg)
\bigg]
  \Bigg\}.
\label{Pzero}
\end{eqnarray}

In the above Eqs.~(\ref{nzero}), (\ref{Ezero}), and (\ref{Pzero}),
$\mbox{arcsh}(x_i)\equiv \ln(x+\sqrt{x^2+1})$ is the inverse hyperbolic sine,
$\mbox{arctan}(x_i)$ and $\mbox{arccot}(x_i)$ are the inverse tangent
and cotangent functions. Please note, the interaction part of
the quark mass scaling, i.e., the second term on the right hand side of
Eq.~(\ref{mass}), simply gives $m_{\mathrm{I}}=D/n_{\mathrm{b}}^z$
at zero temperature.

\section{Properties of strangelets}\label{strangelets}

Thanks to the pioneer works of Witten and Jaffe \textit{et al.}
\cite{Witten1984,Farhi84prd}, we have known a lot about slets. The
properties of slets away from beta equilibrium were also
investigated by mode filling in Ref.~\cite{Schaffner1997} where the
authors checked possible strong and weak hadronic decays (also
multiple hadron decays) and found that slets stable against strong
decays were most likely highly negatively charged. Mode filling
studies are very important to show shell effect \cite{Gilson1993},
but it is difficult for large baryon numbers. Therefore, a
multi-expansion liquid-drop model was developed \cite{Madsen1993}.
Similar studies were done by He \textit{et al.} at finite
temperature \cite{He1996a} and Zhang \textit{et al.} with their
suggested quark mass scaling (QMDTD) \cite{Zhang2003}.
%
In this section, we apply our newly derived quark mass scaling in
Eq.~(\ref{mass}) to investigate the properties of slets.

In relativistic heavy ion experiments, a slet, if formed, has no
time to be in perfect beta equilibrium. We therefore regard it as a
mixture of $u$, $d$ and $s$ quarks. Given three conserved
quantities, i.e., the baryon number $A$, strangeness $S$ and
electric charge $Z$, we have the following equations:
\begin{eqnarray}
A&=&\frac{1}{3}(N_u+N_d+N_s),\label{eq41}\\
Z&=&\frac{2}{3}N_u-\frac{1}{3}N_d-\frac{1}{3}N_s,\label{eq43}\\
S&=&N_s,\label{eq42}
\end{eqnarray}
where $N_u$, $N_d$, $N_s$ are the number of $u$, $d$, and $s$
quarks. For the mechanically stable slets, the internal pressure must
be zero, i.e.
\begin{equation}P=0.\label{eq44}
\end{equation}

We define the charge to baryon number ratio and strangeness fraction
as $f_z = Z/A$ and $f_s=S/A=N_s/A$. Two different linear
combinations of the Eqs.\ (\ref{eq41}) and (\ref{eq43}) give
$N_u=A+Z$ and $N_d+N_s=2A-Z$. We then easily get
\begin{eqnarray}f_z &=& N_u/A-1,\\
f_s &=& 3-(N_u+N_d)/A.
\end{eqnarray}

Because we consider only $0<N_u<3A$ and $0<N_d<3A$, the possible range
for $f_z$ and $f_s$ are $-1\leq f_z \leq 2$ and $-3\leq f_s \leq 3$.
With a view to the relation $f_z+f_s=2-N_d/A$, we have $-1-f_z \leq
f_s \leq 2-f_z$ if $f_z$ is fixed, and we can write $-1-f_s \leq f_z
\leq 2-f_s$ if $f_z$ is given.

\begin{figure}[htbp]
\begin{minipage}[t]{0.48\linewidth}
\centering
\includegraphics[width=7cm,height=7cm]{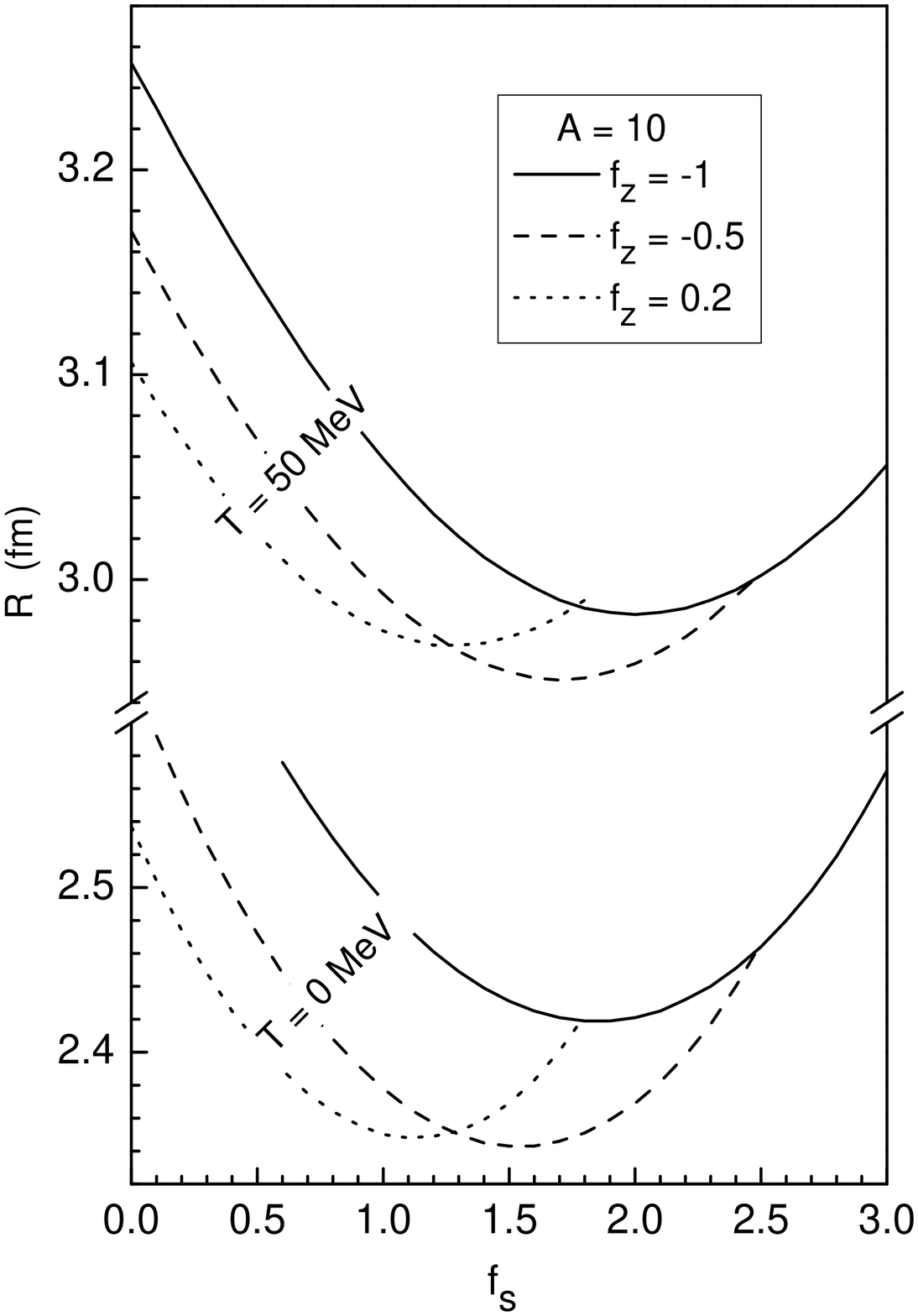}
\caption{
 The mechanically stable radii of a strangelet with baryon number $A=10$
 at temperature $T=0$ and 50 MeV is shown as functions of its strangeness
 fraction $f_s$ for given ratios of charge to baryon number.
 A minimum is reached on each curve.
        }
\label{Rvsfs}
\end{minipage}
\hfill
\begin{minipage}[t]{0.48\linewidth}
\centering
\includegraphics[width=7cm,height=7cm]{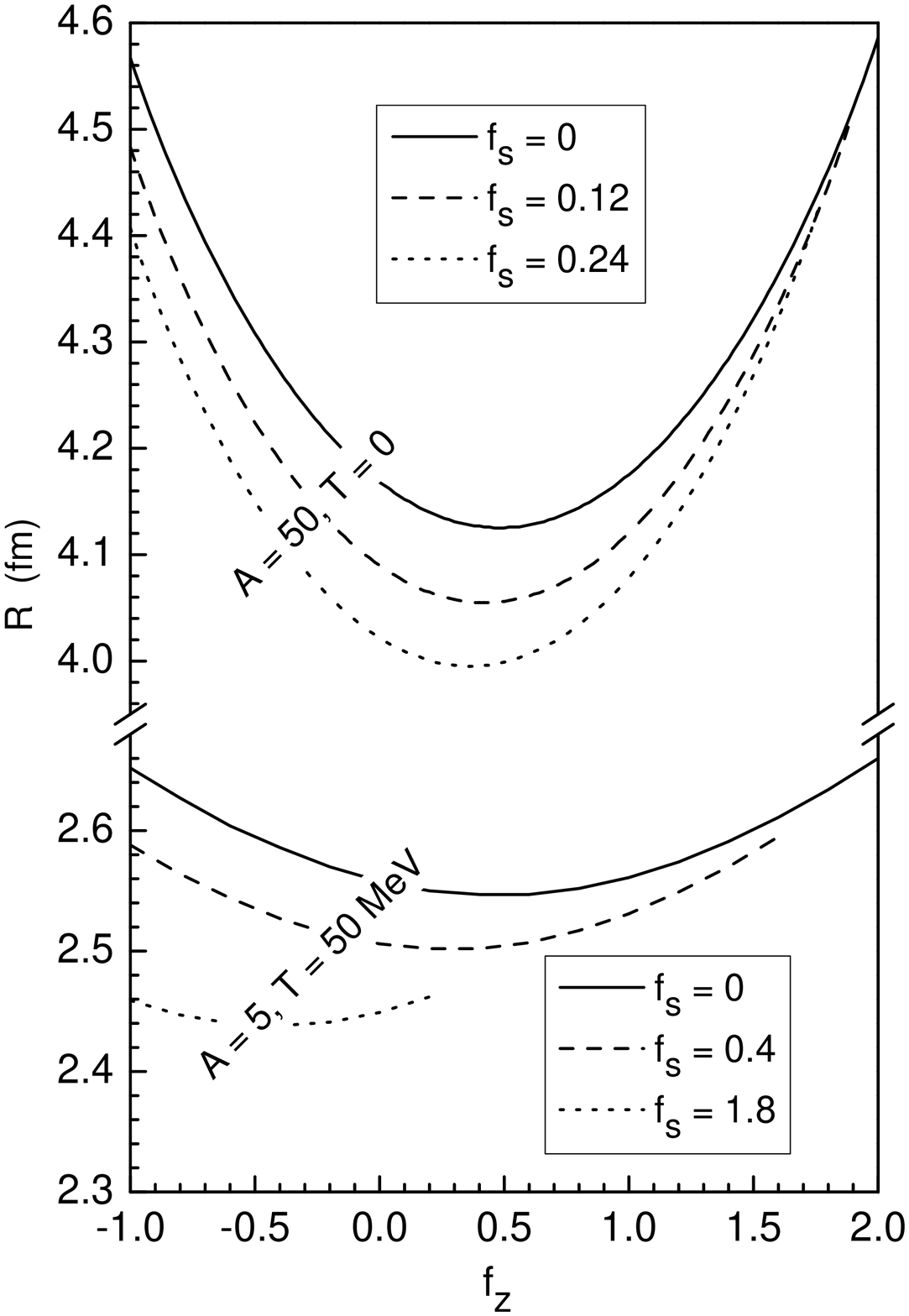}
\caption{
 The radius of a strangelet as functions of the charge to baryon number ratio
 at different given strangeness fraction.
 The up part is for $A=50$ and $T=0$, while the down part
 is for $A=5$ and $T=50$ MeV.
 One can also find a minimum on every curve.
       }
\label{Rvsfz}
\end{minipage}
\end{figure}

For a definite set of values for $A, f_z, f_s$, one can easily get
the quark numbers by $N_u=A(1+f_z), N_d=A(2-f_z-f_s), N_s=Af_s$. If
we give an arbitrary value to the radius $R$, then the slet volume
is $V=(4/3)\pi R^3$, and the density is $n_{\mathrm{b}}=A/V=3A/(4\pi
R^3)$. At a given temperature, the quark density and temperature
dependent mass can be accordingly calculated from Eq.~(\ref{mass}).
The respective chemical potential $\mu_i\ (i=u,d,s)$ can then be
obtained by solving $n_i=3N_i/(4\pi R^3)$ with the $n_i$ expression
in Eq.~(\ref{number}). The pressure and free energy density are
calculated by Eqs.~(\ref{Pexp}) and (\ref{Fexp}), respectively. We
numerically vary $R$ until the pressure becomes zero when the
mechanically stable radius is reached.

At the fixed baryon number $A=10$, we give the slet radius as  a
function of the strangeness fraction $f_s$ at $T=0$ (the down part)
and $T=50$ MeV (the up part) in Fig.~\ref{Rvsfs}. The solid, dashed,
and dotted curves correspond, respectively, to the charge to baryon
number ratio $f_z =-1$, $-0.5$, and 0.2. It is obvious that the
radius is not a monotonic function of strangeness. The position of
the minimum radius depends on electric charge fraction. It is also
natural that the radius increases with increasing temperature.

Similarly in Fig.\ \ref{Rvsfz}, we show the radius of strangelets as
functions of the charge fraction. The variation of the radius with
respect to electric charge is not monotonic at the fixed
strangeness. The curves are also parabolas as in the above Fig.\
\ref{Rvsfs}. Generally, the minimum radius will appear at the middle
position of the region ($-1$, $2-f_s$).

\begin{figure}[htb]
\centering
\includegraphics[width=7cm,height=7cm]{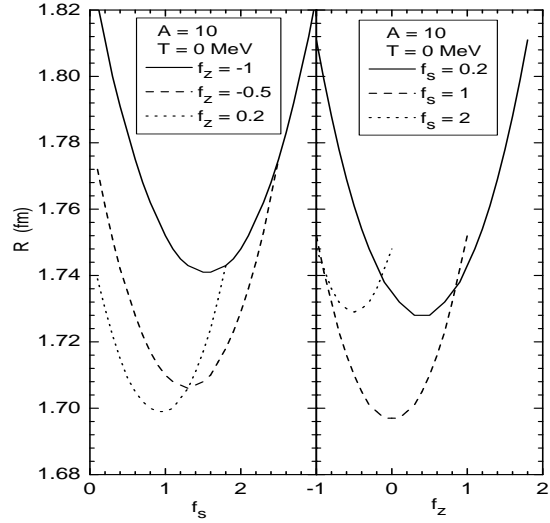}
\caption{
 The strangeness (left panel) and charge (right panel)
 dependence of the strangelet radius in the conventional bag model.
 The relevant parameters are $A=10$, $T=0$ and $B^{1/4}=180$ MeV.
 A minimum is also obviously reached on every curve.
        }
\label{bagfsfz}
\end{figure}

For comparison, we have also done calculations with the conventional
bag model, with the results shown in Fig.\ \ref{bagfsfz}. The left
(right) panel shows the radius with respect to strangeness (charge)
fraction for $A=10$ at zero temperature. One also finds a minimum on
each curve. Therefore, it is safe to conclude that the radius of
slets is not monotonic. Instead, it has a minimum with respect to
the charge and strangeness. This minimum is the smallest radius for
a given pair of baryon number and temperature. However, the smallest
radius does not necessarily correspond to the thermodynamically
stable slets obtained by minimizing the free energy.

\begin{figure}[htb]
\begin{minipage}[t]{0.48\linewidth}
\centering
\includegraphics[width=7cm,height=7cm]{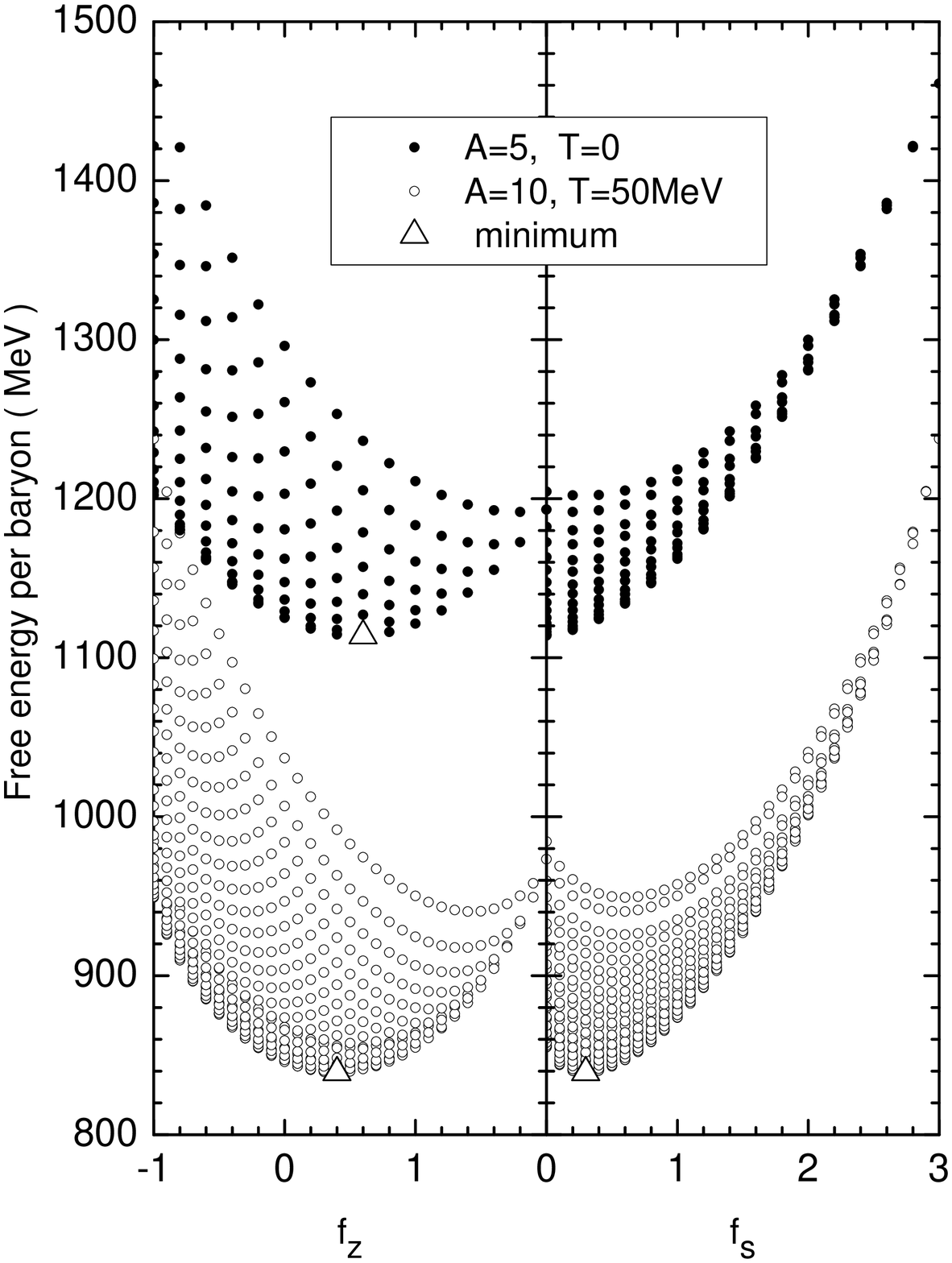}
\caption{
 The free energy of strangelets versus the ratio of charge to
 baryon number (left panel) and the strangeness fraction (right panel) for the
 two cases: $A=5$, $T=0$ and $A=10$, $T=50$.} \label{Fvsfs}
\end{minipage}
\hfill
\begin{minipage}[t]{0.48\linewidth}
\centering
\includegraphics[width=7cm,height=7cm]{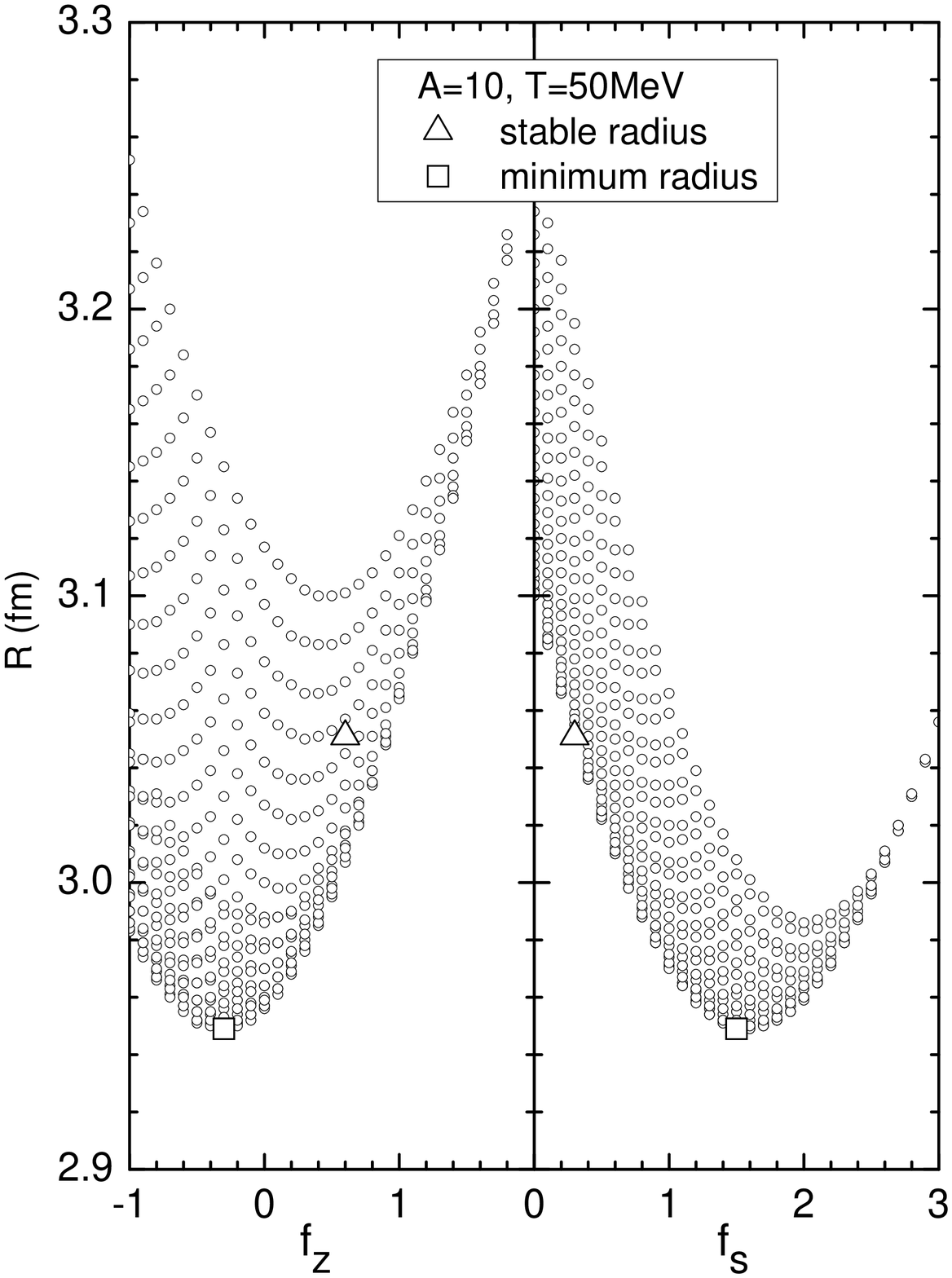}
\caption{
 Comparison of the smallest radius (the square) and the thermodynamically
 stable radius (the triangle) in the $R$-$f_z$ plane (left panel) and in
 the $R$-$f_s$ (right panel) for $A=10$ and $T=50$ MeV.}
\label{radiusA10}
\end{minipage}
\end{figure}

Now we investigate the thermodynamically stable radius of a slet
with given baryon number at fixed temperature. In order to
understand how the stable radius appears analytically, we may fall
back on the fundamental differentiation equality of thermodynamics,
i.e.,
\begin{equation}
dF=SdT-PdV+\sum_i\mu_idN_i.
\end{equation}
 At fixed temperature $T$ and $P=0$, we have
\begin{equation}
dF=\sum_i\mu_idN_i.
\end{equation}
If $A$ and $Z$ are fixed, we obtain naturally,
\begin{equation}\label{fsmini}
\frac{\partial (F/A)}{\partial f_s}=\mu_s-\mu_d.
\end{equation}
 Similarly, with fixed $A$ and $N_s$ we get
\begin{equation}\label{fzmini}
\frac{\partial (F/A)}{\partial f_z}=\mu_u-\mu_d.
\end{equation}
From Eqs.~(\ref{fsmini}) and (\ref{fzmini}), we draw an important conclusion
that the minimum of the free energy per baryon occur when
$\mu_d=\mu_s$ with fixed $A$ and $Z$ in the $F/A$-$f_s$ panel, and
when $\mu_u=\mu_d$ with fixed $A$ and $S$ in the $F/A$-$f_z$ panel.
Therefore, we can get the stablest radius for a slet with given
$A$ and $T$ by requiring the condition
\begin{equation} \label{commonmu}
\mu_u=\mu_d=\mu_s,
\end{equation}
i.e., in this case only one chemical potential is independent. The
only independent one can be determined by solving the equation
\begin{equation} \label{nbR}
\frac{1}{3}(n_u+n_d+n_s)=n_{\mathrm{b}}\ \ \mbox{with}\ \
n_{\mathrm{b}}=\frac{3A}{4\pi R^3}
\end{equation}
for an arbitrary radius. We finally vary the radius $R$
so that the zero pressure condition is satisfied, and accordingly
we obtain the stable radius.

As mentioned in the introduction section, the size of slets
is very important for analyzing their propagation and detection.
Therefore, let's pay special attention to the case of zero temperature
and try to derive an approximate expression for the radius.

Because $\nu_i=\sqrt{\mu_i^2-m_i^2}$, Eq.~(\ref{commonmu}) is equivalent to
\begin{equation}
\sqrt{\nu_u^2+m_u^2}=\sqrt{\nu_d^2+m_d^2}=\sqrt{\nu_s^2+m_s^2}\equiv \mu^*,
\end{equation}
where we have used $\mu^*$ as the common effective chemical
potential. This means $\nu_i=\sqrt{\left.\mu*\right.^2-m_i^2}$, or
$x_i=\sqrt{(\mu^*/m_i)^2-1}$ where $i=u, d,$ or $s$ quarks.
Substituting these into the $n_i$ given in Eq.~(\ref{number}) or
Eq.~(\ref{nzero}) at zero temperature, then substituting into
Eq.~(\ref{nbR}), we obtain an equation which contains the common
chemical potential $\mu^*$ and the radius R. Similarly substitution
into the pressure expression in Eq.~(\ref{Pexp}) or (\ref{Pzero}),
we get another equation of $\mu^*$ and R. The radius is then
obtained by solving the two equations of $\mu^*$ and $R$.

Because the surface and curvature terms can be regarded as a perturbation
to the volume terms, and also because the interaction part of the quark mass
is greater than the quark current masses ($m_{\mathrm{I}}>m_{i0}$),
we can ignore the finite-size effect and iso-spin effect to derive a
first-order approximation for the radius. In this case, we have
$\nu_u=\nu_d=\nu_s\equiv \nu_0$ and $x_u=x_d=x_s\equiv x_0$,
and the zero pressure condition becomes
\begin{equation}
 x_0(2x_0^2-3)\sqrt{x_0^2+1}+3\mbox{arcsh}(x_0)
 -12z \left[ x_0\sqrt{x_0^2+1}-\mbox{arcsh}(x_0) \right].
\end{equation}
The positive solution of this equation is $x_0\approx 1.3278478$
(The trivial solution $x_0=0$ and the non-physical solution
$x_0= -1.3278478$ were discarded).
At the same time, Eqs.~(\ref{nzero}) and (\ref{nbR}) at the same order
approximation gives
$
n_i=\nu_i^3/\pi^2=n_{\mathrm{b}}=3A/(4\pi R^3).
$
Combining this with
$
x_0
=\nu_0/m_{\mathrm{I}}
=(\pi^2n_{\mathrm{b}})^{1/3}/(D/n_{\mathrm{b}}^{1/3})
=(\pi n_{\mathrm{b}})^{2/3}/D,
$
or $n_{\mathrm{b}}=(x_0D)^{3/2}/\pi$,
we immediately have 
\begin{equation} \label{Rapprox}
R=\frac{(3/4)^{1/3}}{\sqrt{x_0D}} A^{1/3}\equiv r_{\mathrm{slet}} A^{1/3}.
\end{equation}
For the chosen value $D=(156\ \mbox{MeV})^2$, the reduced slet radius is
\begin{equation}
r_{\mathrm{slet}}=R/A^{1/3}=\left.(3/4)^{1/3}\right/\sqrt{x_0D}=0.9973\ \mbox{fm}.
\end{equation}
This value is smaller than that of normal nuclei,
but bigger than the recent value ($\sim$ 0.94 fm) \cite{Wufei2007jpg}
from the conventional bag model calculations.
Equation (\ref{Rapprox}) shows that the slet radius is inversely proportional
to the square root of the confinement parameter $D$. If one uses a bigger
D value, then the slet radius can be very small. In that case, however, SQM
will not be absolutely stable.

It should be emphasized that the expression for the slet radius in
Eq.~(\ref{Rapprox}) is only the lowest order approximation because
it was obtained by ignoring the finite size effect and isospin
dependence (quark mass difference). The actual size is a a little
bit bigger, from 1 fm (for large baryon numbers) to about 1.2 fm
(for small baryon numbers).
So Eq.~(\ref{Rapprox}) is only accurate for slets with large baryon numbers.
In the following we continue to present the numerical results.

\begin{figure}[htb]
\centering
\includegraphics[width=8cm,height=8cm]{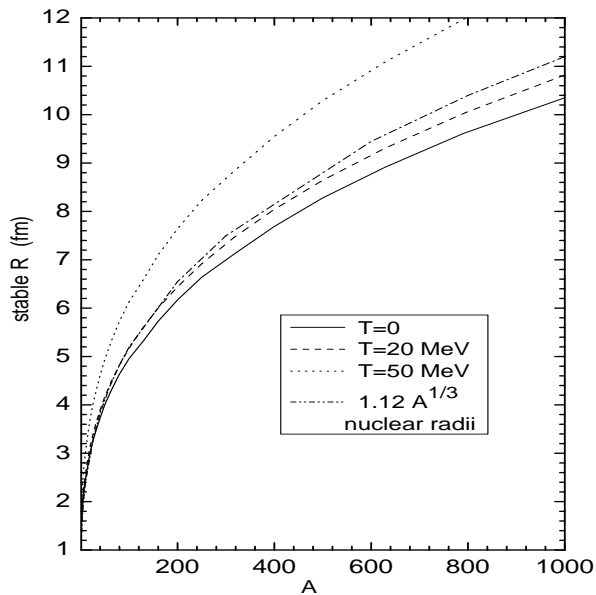}
\caption{
 The baryon number dependence of the strangelet radius at
 temperature $T=0$ (solid curve), 20 MeV (dashed curve), and 50 MeV (dotted curve).
 The dash-dotted curve is for the ordinary nuclei radius $R=1.12A^{1/3}$ fm.
        }
\label{AstableR}
\end{figure}

We show all possible slets with positive strangeness in
Fig.~\ref{Fvsfs}. The full dots are for $A=5$ and $T=0$, while the
open circles are for $A=10$ and $T=50$ MeV. The left (right) panel
gives the distribution of the free energy per baryon on the $f_z$
($f_s$) axis. The triangle denotes the minimum of the free energy
per baryon, i.e., the most stable slets. For a given pair of $A$ and
$T$, therefore, we have two special values for the radius of a slet
via varying the strangeness fraction $f_s$ and the charge to baryon
number ratio $f_z$: the first one is obtained by minimizing the
radius itself, and thus is the smallest radius, while the second one
is obtained by minimizing the free energy of the system, and is thus
the thermodynamically stable radius. The obvious difference between
the smallest radius (square) and the stable radius (triangle) is
compared in Fig.\ \ref{radiusA10}, where the smallest radius is 2.95
fm for $A=10$ and $T=50$~MeV while the stable radius is 3.05 fm.

The baryon number dependence of the stable slet radius is shown in
Fig.~\ref{AstableR} at temperature $T=0$ (solid line), 20 MeV
(dashed line), and 50 MeV (dotted line). To compare, the normal
nuclear radii have also been shown with a dash-dotted line. We see
that the radii of slets are comparable with those of ordinary
nuclei.

\begin{figure}[htb]
\begin{minipage}[t]{0.48\linewidth}
\centering
\includegraphics[width=8cm,height=8cm]{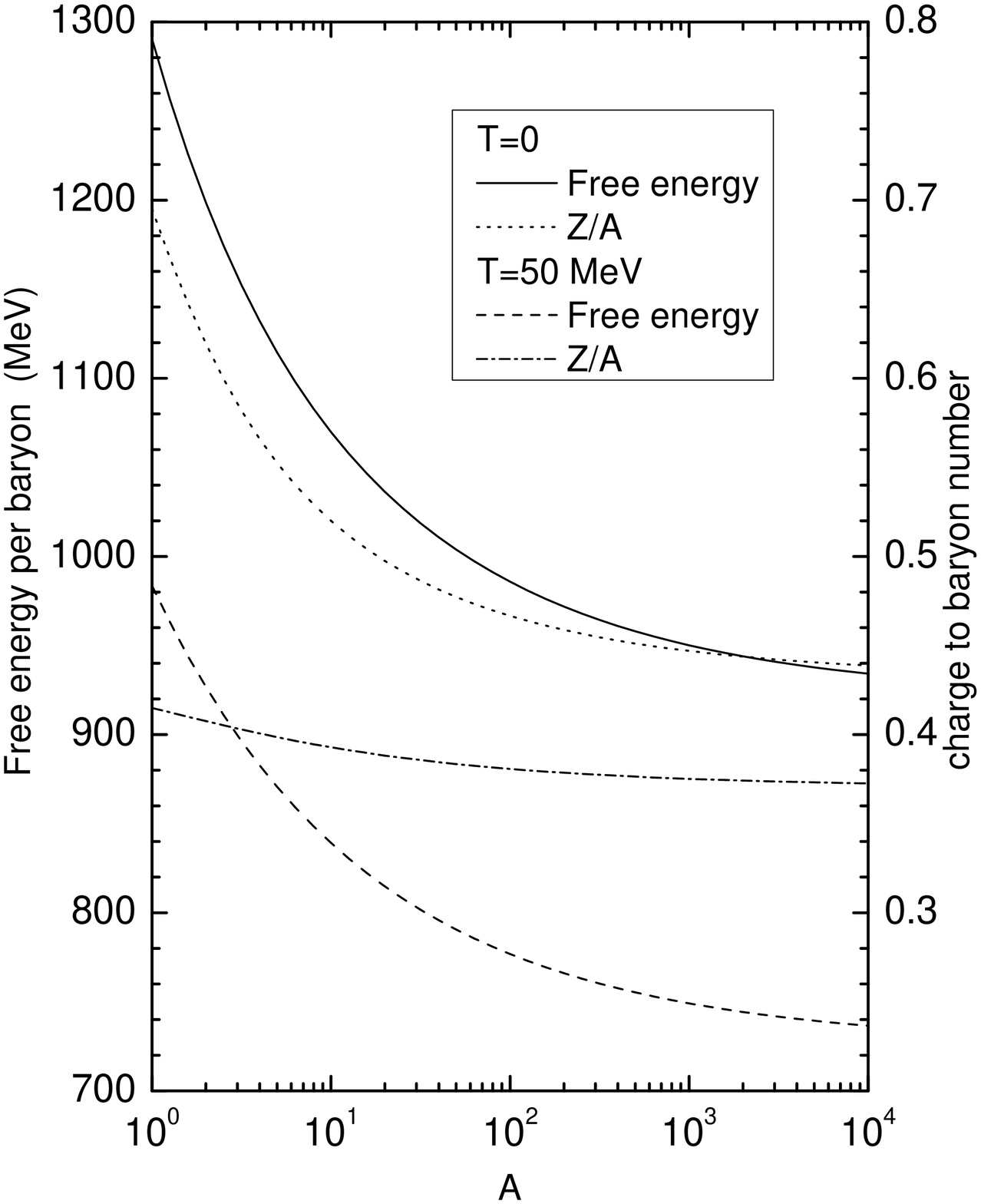}
\caption{
 The free energy per baryon (left axis) and the charge to
 baryon number $Z/A$ (right axis) of stable strangelets
 vary with the baryon number $A$. They are generally decreasing functions.
 With increasing temperature, the decreasing speed goes down, and
 $Z/A$ is very flat at high temperature.
         }
\label{AfnT50}
\end{minipage}
\hfill
\begin{minipage}[t]{0.48\linewidth}
\centering
\includegraphics[width=8cm,height=8cm]{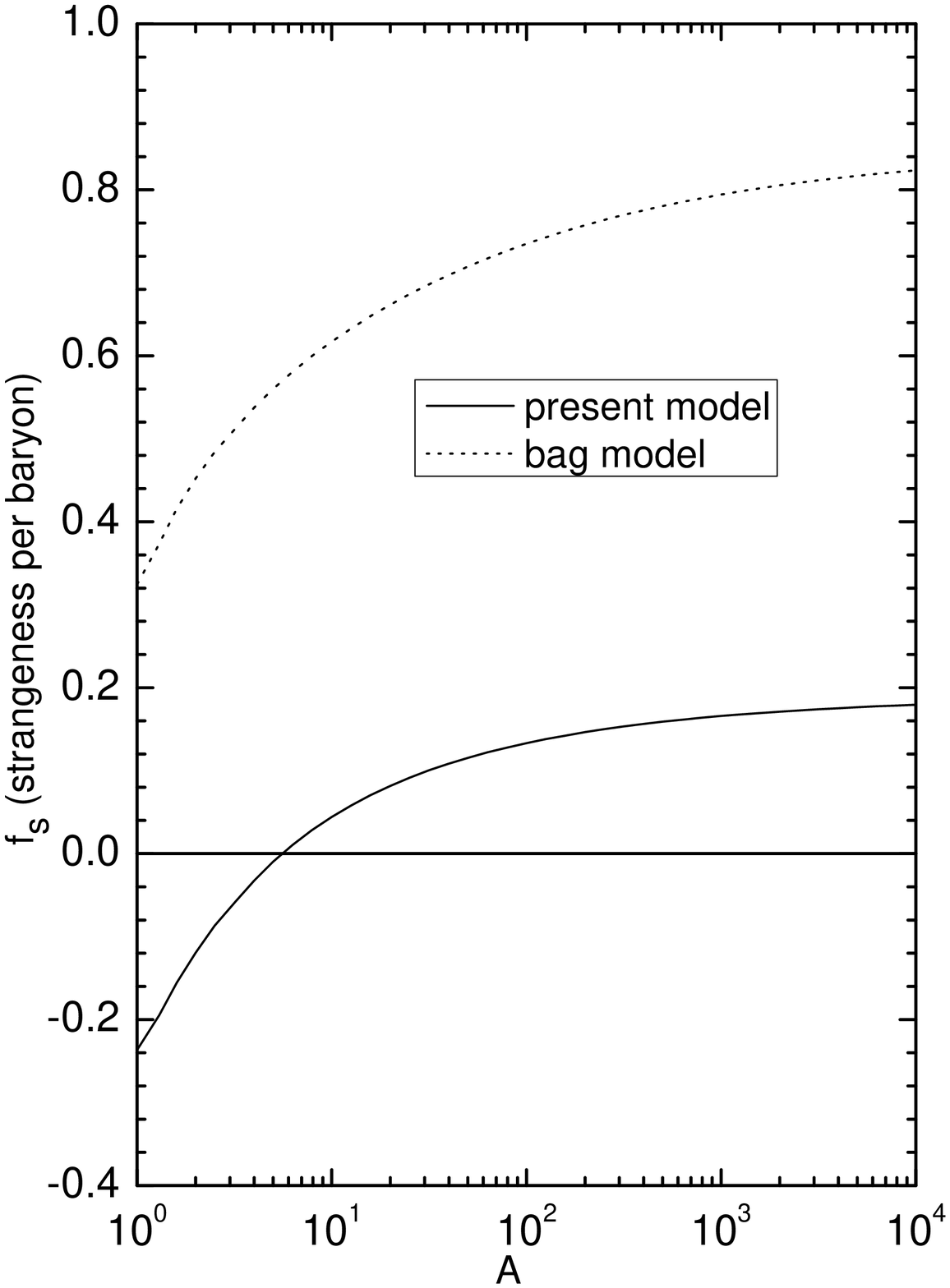}
\caption{
 The strangeness per baryon for stable strangelets
 as a function of the baryon number at zero temperature.
 For very small baryon numbers, the strangeness fraction in the present
 model (solid curve) becomes negative, while that in the bag model (dotted curve)
 is always positive.}
\label{negsness}
\end{minipage}
\end{figure}

Fig.\ \ref{AfnT50} gives the free energy per baryon and the charge to
baryon number ratio of stable slets as a function of the
baryon number. It is seen that the free energy per baryon decreases
with both increasing temperature and baryon number. When the
temperature is as high as up to 50 MeV, the dash-dotted line for
charge to baryon number becomes very flat. The smaller the baryon
number, the more important the influence of temperature on the
charge to baryon number.

In Fig.\ \ref{negsness}, we show the strangeness fraction $f_s$,
with a solid line, as a function of the baryon number at zero
temperature. The striking feature is that the strangeness fraction
becomes negative for very small baryon numbers. In fact, this
feature has already been seen in the up part of the right panel in
Fig.\ \ref{Fvsfs} where the minimum of the free energy was not
reached at positive strangeness. For comparison, we also give, in
the same figure with a dotted line, the results from the
conventional bag model in which the quark masses are constant and
the strangeness fraction is always positive. In the present model,
however, the quark masses are density and temperature dependent.
Therefore, the strangeness fraction becomes negative for very small
baryon numbers, and the charge to baryon number ratio becomes bigger
than that in the bag model.

\section{Summary} \label{sum}

  We have studied the properties of strangelets in a new quark mass
scaling without imposing beta equilibrium. It is found that the
radius of strangelets is not a monotonic increasing or decreasing
function of either electric charge or strangeness. By varying the
strangeness and charge for a given pair of baryon number and
temperature, we have calculated the smallest radius and
thermodynamically stable radius, and have shown that they are
generally different. The stable radii of strangelets can be
calculated approximately by the law $R=r_{\mathrm{slet}} A^{1/3}$.
If SQM is absolutely stable, the reduced radius of strangelets is
$r_{\mathrm{slet}}=(3/4)^{1/3}/\sqrt{x_0D}\sim 1$ fm. The smallest
radius appears always at positive strangeness. However, the stable
strangelets could have negative strangeness fraction for very small
baryon numbers. This means that very small strangelets may contain
anti-strangeness.

The present study is still model-dependent. Many problems, e.g., how
the anti-strangeness content influences the stability and
detectability of strangelets in heavy ion experiments, to what
extent the behavior of the radius can be used to analyze the
production and detection of strangelets, etc, need to be further
explored.

\Large\begin{flushleft} {\bf
Acknowledgements}\end{flushleft}\normalsize

The authors thank support from the Natural Science Foundation of
China (10675137, 10375074, and 90203004), and KJCX3-SYW-N2.

\end{document}